\documentclass[8pt]{article}
\renewcommand{\baselinestretch}{1.5}

\usepackage{graphicx}
\usepackage{bm}
\usepackage{color}
\usepackage{textcase, url}
\usepackage[colorlinks=true, linkcolor=black, urlcolor=black, citecolor=black]{hyperref}
\usepackage{amsmath,amssymb}
\usepackage{mathtools}
\usepackage{lineno}
\usepackage{tabularx}
\usepackage{wrapfig}
\usepackage{booktabs}
\usepackage{lettrine}
\usepackage{lipsum}
\usepackage[a4paper,bottom=20mm,right=20mm,left=20mm,top=20mm]{geometry}
\usepackage{float}
\usepackage{indentfirst}
\usepackage[separate-uncertainty = true,multi-part-units=single]{siunitx}
\usepackage[margin=0mm]{caption}
\usepackage{subcaption}
\usepackage[utf8]{inputenc}

\usepackage[
    backend=biber,
    style=nature,
    sorting=none,
    doi=false,
    url=false,
    date=year
]{biblatex}
\AtEveryBibitem{
  \clearfield{issn}  
  \clearfield{month} 
  \clearfield{issue} 
}
\DeclareFieldFormat[misc]{title}{#1} 

\DeclareBibliographyDriver{misc}{%
  \usebibmacro{author/editor+others}%
  \newunit\newblock
  \printfield{title}%
  \newunit\newblock
  \printfield{eprint}%
  \newunit\newblock
  \printfield{howpublished}%
  \newunit\newblock
  \printfield{note}%
  \newunit\newblock
  \printtext[parens]{\printfield{year}}
  \finentry
}

\begin{document}
\setlength{\abovedisplayskip}{8pt} 
\setlength{\belowdisplayskip}{8pt} 

\begin{center}
    \textsf{\textbf{\LARGE Electronic interferometry with ultrashort plasmonic pulses} } 
    \\\vspace{2mm}
    
    {\small
        Seddik Ouacel$^{1}$,
        Lucas Mazzella$^{1}$,
        Thomas Kloss$^{1}$,
        Matteo Aluffi$^{1}$,
        Thomas Vasselon$^{1}$,
        Hermann Edlbauer$^{1}$,
        Junliang Wang$^{1}$,
        Clement Geffroy$^{1}$,
        Jashwanth Shaju$^{1}$,
        Arne Ludwig$^{10}$,
		Andreas D. Wieck$^{10}$,
        Michihisa Yamamoto$^{4,5}$,
        David Pomaranski$^{5}$,
        Shintaro Takada$^{7,8,9}$,
        Nobu-Hisa Kaneko$^{6}$, 
        Giorgos Georgiou$^{2}$,
        Xavier Waintal$^{3}$,
        Matias Urdampilleta$^{1}$,
        Hermann Sellier$^{1}$\&
		Christopher B\"auerle$^{1,\star}$
	}
\end{center}
\vspace{-2mm}
\def\einr{2mm}
\def\spazi{-1.7mm}
{\small
\begin{nolinenumbers}
\-\hspace{\einr}$^1$ Universit\'e Grenoble Alpes,       CNRS, Grenoble INP, Institut N\'eel, F-38000 Grenoble, France\\
\-\hspace{\einr}$^2$ James Watt School of           
   Engineering, Electronics and Nanoscale Engineering, University of Glasgow, Glasgow G12 8QQ, United Kingdom\\
\-\hspace{\einr}$^3$ Universit\'e Grenoble Alpes, CEA, INAC-Pheliqs, F-38000 Grenoble, France\\
\-\hspace{\einr}$^4$ Center for Emergent Matter Science (CEMS), RIKEN, Saitama, Japan\\
\-\hspace{\einr}$^5$ Quantum-Phase Electronics Center and Department of Applied Physics, The University of Tokyo, Tokyo, Japan\\
\-\hspace{\einr}$^6$ National Institute of Advanced Industrial Science and Technology (AIST), National Metrology Institute of Japan\vspace{\spazi}\\
\-\hspace{\einr}\-\hspace{3mm}(NMIJ), Tsukuba, Ibaraki, Japan\\
\-\hspace{\einr}$^7$ Department of Physics, Graduate School of Science, Osaka University, Toyonaka 560-0043, Japan\\
\-\hspace{\einr}$^8$ Institute for Open and Transdisciplinary Research Initiatives, Osaka University, Suita 560-8531, Japan\\
\-\hspace{\einr}$^9$ Center for Quantum Information and Quantum Biology (QIQB), Osaka University, Osaka 565-0871, Japan\\
\-\hspace{\einr}$^{10}$Lehrstuhl für Angewandte Festkörperphysik,
Ruhr-Universität Bochum, Bochum, Germany\\
\-\hspace{\einr}$^\ast$ To whom correspondence should be addressed;\\


	\-\hspace{\einr}$^\star$ corresponding author:
	\href{mailto:christopher.bauerle@neel.cnrs.fr}{christopher.bauerle@neel.cnrs.fr}

	\vspace{2mm}
\end{nolinenumbers}
}

\def\baselinestretch{1.5}\selectfont


{
\bf 
Electronic flying qubits offer an interesting alternative to photonic qubits: electrons propagate slower, hence easier to control in real time, and Coulomb interaction enables direct entanglement between different qubits. 
Although their coherence time is limited, flying electrons in the form of picosecond plasmonic pulses could be competitive in terms of the number of achievable coherent operations.
The key challenge in achieving  this critical milestone
is the development of a new technology capable of injecting `on-demand' single-electron wavepackets into quantum devices, with temporal durations comparable to or shorter than the device dimensions.
Here, we take a significant step towards achieving this regime in a quantum nanoelectronic system
by injecting ultrashort single-electron
plasmonic pulses into a 14-micrometer-long Mach-Zehnder interferometer. 
Our results establish that quantum coherence is robust under the on-demand injection of ultrashort plasmonic pulses, as evidenced by the observation of coherent oscillations in the single-electron regime.
Building on this, our results demonstrate for the first time the existence of a new 
"non-adiabatic" regime that is prominent  at high frequencies.
This breakthrough highlights the potential of flying qubits as a promising alternative to localised qubit architectures, offering advantages such as a reduced hardware footprint, enhanced connectivity, and scalability for quantum information processing.}


\begin{refsection}
Solid-state systems, presently considered for quantum computation, are built from localised two-level systems. 
Prime examples are superconducting qubits or semiconducting quantum dots \cite{Kjaergaard2020,Burkard2023}. Being localised, they require a fixed amount of hardware per qubit. 
Conversely, flying qubits are the only existing quantum technology platform that uses propagating particles.
They represent an interesting quantum architecture, as they naturally enable the implementation of quantum interconnects. 
Currently, flying qubits are associated with photons due to their highly coherent nature, on-demand generation, and inherent scalability \cite{Slussarenko2019}.
Photons, however, travel so fast that in-flight dynamical manipulation is impossible, and their trajectories need to be set in advance. 
Moreover, they do not interact directly with each other, making photon entanglement challenging, and the ‘all-linear’ quantum optics approach \cite{Romero2024} requires post-selection methods.
As a result of the very weak photon interaction, a large number of Mach-Zehnder interferometers must be implemented to construct a single two-qubit gate \cite{Maring2024}.
This inevitably leads to a tremendous increase in hardware overhead.


Despite their much shorter coherence time compared to photons, quantum nanoelectronic circuits have seen enormous progress over the last 10 years.
This advancement was driven by the development of on-demand single-electron sources capable of generating single-electron wavepackets with high fidelity \cite{Fève2007,Blumenthal2007,Dubois2013,Wang2022}. 
Moreover, Coulomb interactions between two individual co- \cite{Wang2023} and counter-propagating electrons \cite{Ubbelohde2023,Fletcher2023} have been successfully demonstrated.
Such progress marks a significant milestone, paving the way for the entanglement of multiple flying electron qubits in the future \cite{Bäuerle2018,Edlbauer2022,Bocquillon2014}.

The most convenient method to generate a single-electron excitation is by applying a short voltage pulse to the Ohmic contact of a two-dimensional electron gas (2DEG). This creates a single electron-excitation in the form of a plasmonic pulse \cite{Roussely2018}.
Recent experiments have demonstrated the coherent manipulation of single-electron plasmonic pulses in the form of Levitons within a Mach-Zehnder interferometer (MZI) implemented in graphene \cite{Assouline2023}, highlighting their potential for quantum information processing.

The next milestone towards developing a competitive quantum architecture for flying electron qubits is to reach a regime where the wavepacket's
width is significantly shorter than the quantum device.
Such achievement would allow multiple flying qubits to be accommodated within a single quantum processing unit and would enable the implementation of a large number of gate operations during their flight \cite{Pomaranski2024}
Moreover, as the wavepacket width decreases, it inevitably becomes comparable to or shorter than the characteristic timescales of the interferometer. 
This marks the transition into the non-adiabatic regime where dynamical effects are expected to play an important role \cite{Gaury2014}.

In this work, we demonstrate for the first time quantum coherence of ultrashort electron wavepackets in the non-adiabatic regime in an electronic Mach-Zehnder interferometer.
After describing the working principle of the device in the DC regime, we present a detailed characterisation of its nonlinear behavior. This nonlinearity is harnessed to investigate the frequency response of the Mach-Zehnder interferometer via quantum rectification \cite{Rossignol2018}. Through these measurements, we demonstrate the onset of the non-adiabatic regime for frequencies starting around 1 GHz. Finally, we utilize these findings to establish the presence of non-adiabatic coherent current oscillations driven by ultrashort plasmonic pulses of 40\,ps duration.

\newpage
\textbf{Electronic Mach-Zehnder interferometer device:} The cornerstone of a flying qubit platform is the Mach-Zehnder interferometer (MZI). 
Here, individual particles are injected on demand, put in a superposition of states at the initial beam splitter, guided through the interferometer, and ultimately, the quantum superposition can be discerned at the two output detectors.
Throughout the particle's trajectory, quantum manipulations can be implemented by electrical gate operations.

\begin{figure*}[h!]
\centering
\includegraphics[scale=1]{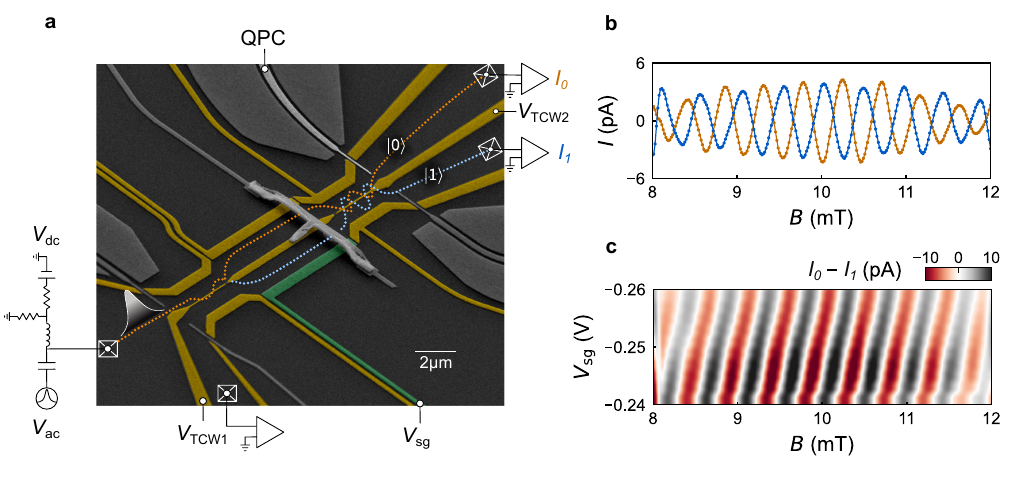}
\caption{
\textbf{Electronic Mach-Zehnder interferometer.} 
\textbf{a.} 
Scanning electron micrograph of the electronic Mach-Zehnder interferometer (MZI) device. 
The electrostatic gates highlighted in color define the electron trajectories indicated by the dotted lines.
Electrons are injected into the interferometer through the left Ohmic contact (crossed white square). 
The output current $I_0$ ($I_1$), corresponding to the transmitted current in the upper (lower) electron waveguide, is measured using a Lock-In amplifier. 
\textbf{b.} Coherent anti-phase oscillations of the transmitted current $I_0$ and $I_1$ when applying a DC bias where a smooth background has been subtracted.
\textbf{c.} 
Coherent oscillations of $I_0$ - $I_1$ (total oscillating component of the transmitted current) as a function of the magnetic field and the side gate voltage $V_{\rm{sg}}$.  
}
\label{fig:fig1}
\end{figure*}

In our system, the qubit states are represented by the presence of an electron in the upper $|0\rangle$ or lower $|1\rangle$ arm of the electronic Mach-Zehnder interferometer of a total length of 14 $\mu$m, as depicted in Fig.\,\ref{fig:fig1}a.
%
%
To create a superposition between the two states, a tunnel-coupled wire (TCW) of a length of 2\,$\mu$m is employed that acts as an electronic beam splitter. 
In this device, the electronic waveguides are brought into close proximity, enabling quantum tunneling of the injected wavepackets  between them. 
The tunneling is controlled via the voltage applied to the middle gate of the tunnel-coupled wire $V_{\rm{TCW}}$, providing full electrical control of the beam splitter.
The wavepackets are then allowed to propagate through an Aharonov-Bohm ring, where a phase difference between the upper and the lower arm can be induced by varying the magnetic flux $\phi$ enclosed by the two paths. 
Alternatively, the phase can be electrically controlled by applying a voltage $V_{\rm{sg}}$ to the side gates \cite{Takada2015}. 
A second beam splitter is placed at the end of the ring to enable interference between the two wavepackets, thereby effectively implementing an electronic Mach-Zehnder interferometer
\cite{Yamamoto2012, Takada2015}.

To investigate the coherent properties of our device, we adopt the following approach. 
Electrons are injected into the electronic MZI by applying a constant bias voltage $V_{\rm{dc}}$ to the injection Ohmic contact, as indicated by the left white square in Fig.\,\ref{fig:fig1}a. 
The transmitted output currents $I_0$ and $I_1$ are then measured as a function of the applied magnetic field $B$. 
When the device is properly tuned into a two-path interferometer \cite{Takada2015}, anti-phase oscillations in the currents are observed at the two outputs, as shown in Fig.\ref{fig:fig1}b.
Here we plot the oscillating component $I$ where a smooth background current ($\approx$ 1 nA) has been subtracted. 
We measure a magnetic field periodicity of the current oscillations of $\Delta B$ =  0.5\,mT, corresponding to a surface area of $S$ = 8.2 $\mu$m$^2$, which is consistent with the device geometry.
Varying the voltage $V_{\rm{sg}}$ of the side gate highlighted in green in Fig.\,\ref{fig:fig1}a, a controlled phase shift between the propagating electrons through the lower arm with respect to the upper one can be achieved. 
In our case a voltage variation of $\Delta V_{\rm{sg}} \approx 20$\,mV is sufficient to induce a phase shift of 2$\pi$ (see Fig.\,\ref{fig:fig1}c).
Demonstrating interference with ultrashort voltage pulses requires a careful understanding of the device's frequency response.
When the MZI is driven by a purely sinusoidal signal, the net average current is theoretically expected to be zero. 
However, as shown in Figs.\,\ref{fig:fig3}a and c, this is not the case. 
Instead, we observe current rectification, attributed to a nonlinearity in the device. 
In the following section, we provide a detailed characterisation of this nonlinearity, which is subsequently utilized to investigate the frequency response of the MZI.

\begin{figure*}[h!]
\centering
\includegraphics[scale=0.9]{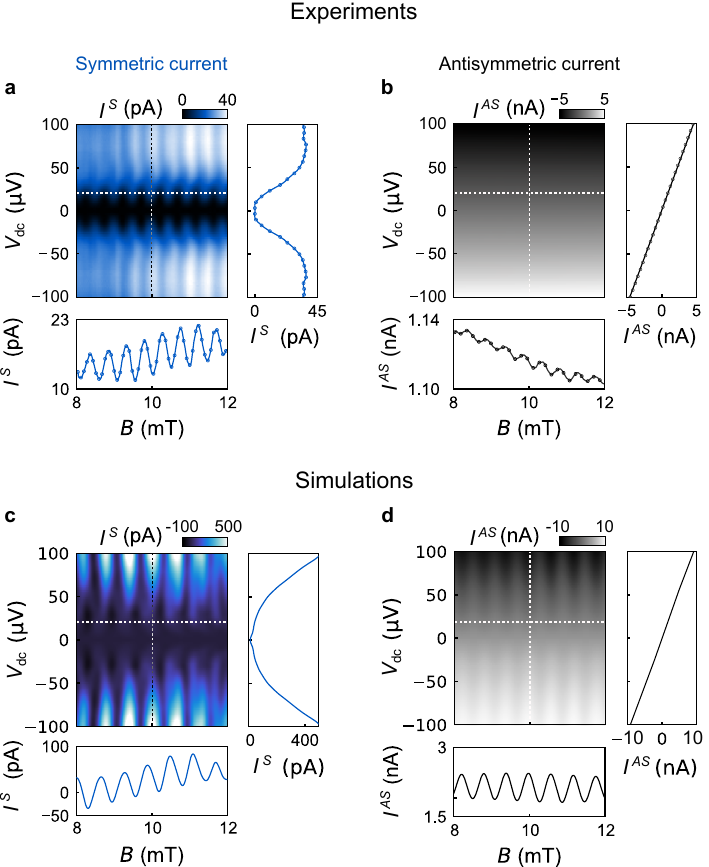}
\caption{\textbf{Nonlinearity of the electronic Mach-Zehnder interferometer}. 
The DC $I$-$V$ curve is decomposed into its symmetric $(I^S)$ and antisymmetric $(I^{AS})$ components with respect to the bias voltage $V$.   
\textbf{a.} Density plot of the symmetric component of the current as a function of the DC bias voltage $V_{dc}$ and magnetic field $B$.
The vertical line cut, shows the non-linear $I$-$V$ characteristic at a magnetic field of $B$ = 10~mT. 
The horizontal line cut shows AB  oscillations at a bias voltage of $V=25$~µV. \textbf{b.} Same as \textbf{a}, but for the antisymmetric component of the current. 
The $I$-$V$ characteristic exhibits primarily  a linear behaviour. The applied bias voltage is corrected to account for the effective reduction of the bias due to the electronic circuit (Supplementary Note, section 1.1).
\textbf{c, d.} Quantum transport simulations of both the symmetric and antisymmetric components of the current, analogous to \textbf{a} and \textbf{b}.
}
\label{fig:fig2}
\end{figure*}

\textbf{Nonlinearity of the device revealed in DC measurements:}
We begin by investigating the source of the nonlinearity responsible for the rectification, and thoroughly characterising the device's DC response.
Nonlinear effects in mesoscopic transport systems have been studied in the past,
with particular focus on  quantum point contacts \cite{Kouwenhoven1989} and Aharonov-Bohm (AB) rings \cite{Angers2007,Leturcq2006}.
In prior research, coherent oscillations in the nonlinear conductance of two-terminal AB rings have been observed, with different proposed origins: in \cite{Angers2007}, the nonlinearity arose from spatial inversion asymmetry while its magnetic field asymmetry was attributed to electron-electron interactions, whereas in \cite{Leturcq2006}, the nonlinearity was suggested to originate from bias-dependent transmission, though its precise microscopic origin remained unclear.

In our system, the nonlinearity originates from the tunnel-coupled wires at the entrance and exit of  the electronic Mach-Zehnder interferometer. 
This is evidenced by experimental measurements of its nonlinear $I$-$V$ curve and further corroborated by numerical simulations.
To illustrate the device's nonlinearity, we decompose the  output current into its symmetric, $I^{S} (V,B) =
 \big (I(V,B) + I(-V,B) \big )$\,/\,2,
and antisymmetric, $I^{AS} (V,B)  =
 \big ( I(V,B) - I(-V,B) \big )$\,/\,2,
components. 
The symmetric current is indeed non-zero, as shown in the vertical line cut of the right panel in Fig.\,\ref{fig:fig2}a, with nonlinearities appearing for bias voltages as low as 25\,$\mu $V.
In contrast, the antisymmetric current is dominated by a linear response, 
as seen in the vertical line cut of the right panel in Fig.\,\ref{fig:fig2}b. 
To further emphasize the contribution of the nonlinearity towards the coherent oscillations, we analyze the magnetic field dependence. 
For the symmetric current, AB oscillations with minimal background current are observed (horizontal line cut of the bottom panel in Fig.\,\ref{fig:fig2}a). 
On the contrary, the antisymmetric current shows 
AB oscillations superimposed on a significant background current (horizontal line cut of the bottom panel in Fig.\,\ref{fig:fig2}b).
These oscillations are the characteristic AB oscillations measured in a linear system.
To further confirm that the nonlinearity primarily originates from the tunnel-coupled wire, we conducted a separate investigation focusing exclusively on this component of our MZI. 
We observed a similar nonlinearity to that seen in the entire MZI, reinforcing our hypothesis (for details see Supplementary Note, section 1.6).
%

Our experimental findings are supported by state-of-the-art quantum transport simulations (Fig.~\ref{fig:fig2}c, d). These simulations combine detailed electrostatic potential simulations \cite{Chatzikyriakou2022} with transport calculations using the Kwant software \cite{Groth2014,Bautze}. 
The electrostatic calculations account for the precise geometric configurations of the surface gates and the properties of the GaAs/AlGaAs heterostructure (see Supplementary Note, section~2.1 for details). 
While the simulations reproduce the main features of our measurements in a semi-quantitative manner, some discrepancies persist. 
These differences likely arise from the simulations not accounting for disorder in the electrostatic potential caused by dopants or for decoherence phenomena.
Nonetheless, our simulations indicate that the observed nonlinearity—and the resulting rectification in our system—originates from the energy-dependent transmission of the tunnel-coupled wire (see Supplementary Note, Fig.~S8 for details).

\textbf{Frequency response of the MZI.}
Having characterised
our electronic MZI in the DC regime we now investigate the response of our device under sinusoidal drive at variable frequencies.
At low frequencies, the sinusoidal signal is slow enough and the system adjusts instantaneously to the external drive - known as the adiabatic regime.
In Fig.\,\ref{fig:fig3}a, we show the amplitude of the coherent current oscillations, $I_{\rm{AB}}$, defined as the maximum amplitude of $I_0 - I_1$, extracted from the fast Fourier transform, as a function of frequency and drive amplitude $V_{\rm{ac}}$.
We observe that for frequencies below 100 MHz, the voltage dependence of the coherent oscillations is independent on frequency and shows the same evolution as a function of  drive amplitude.
To show that the frequency response at low frequencies ($\le$ $100$~MHz) can be described in the adiabatic limit, we reconstruct the  
oscillating component $I_{AB}$, 
generated by a sinusoidal signal directly from the raw DC data, using the following formula 
\begin{equation}
    I_{\rm{sin}}(V,\,B)=\\
    \frac{1}{T}\int I_{\rm{dc}}\big (V(t),\,B\big )\,dt
    \label{eq:I_rect}
\end{equation}
where $I_{\rm{sin}}$ is the DC rectified current induced by a sinusoidal drive $V(t)=V_{\rm{ac}}\sin(\omega t)$. 
The evolution of the amplitude of the coherent oscillations $I_{\rm{AB}}$ passes through a maximum and saturates at high bias.
The overall shape of the evolution is well captured by the DC reconstruction (gray continuous line in Fig.\,\ref{fig:fig3}),
as well as by the Floquet simulation (see Supplementary Note, Section 2.4). 
In these simulations, the exact position of the maximum with respect to the bias voltage depends on the microscopic parameters of the device.
We observe that all experimental data at low frequency ($\leq$ 100 MHz) follow the adiabatic limit.
%
\begin{figure*}[h!]
\centering
\includegraphics[scale=0.85]{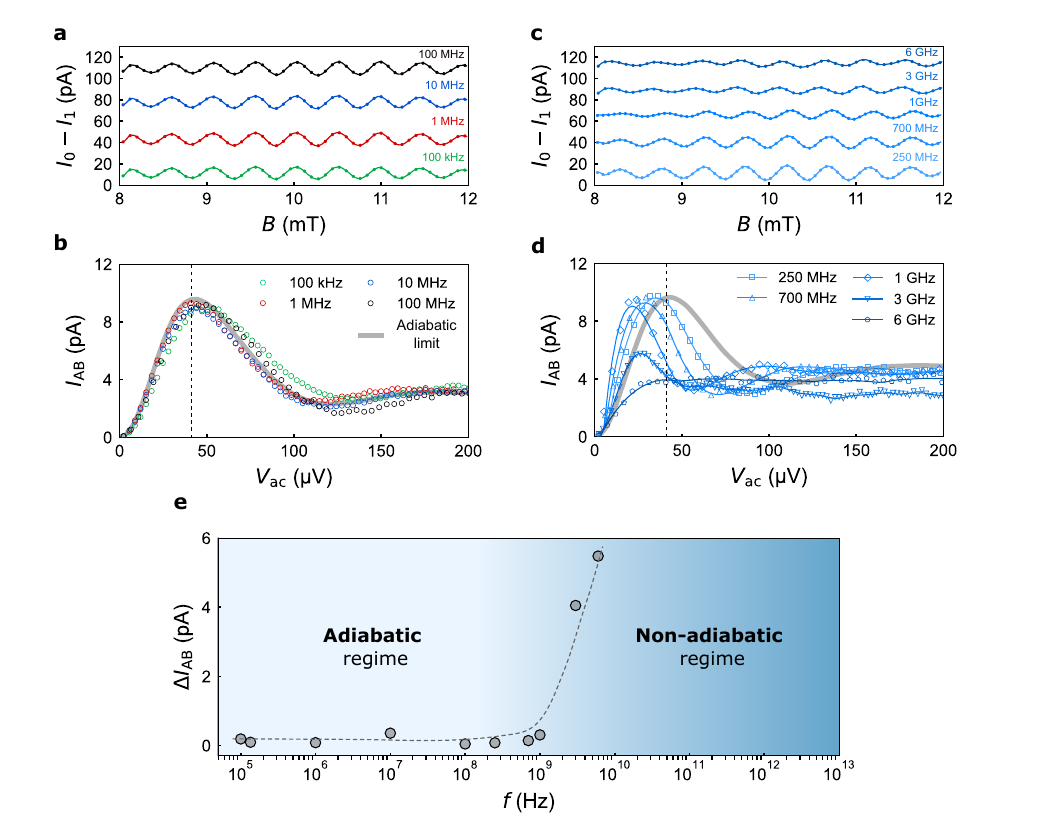}
\caption{
\textbf{Frequency response of the MZI under sinusoidal drive.}
\textbf{a, c.} AB oscillations of the current difference $I_0 - I_1$ for frequencies ranging from 100 kHz to 6 GHz. Experimental data are vertically offset for clarity.
\textbf{b.} Amplitude of the AB oscillations $I_{\rm{AB}}$ versus sinusoidal drive amplitude $V_{\rm{ac}}$ for frequencies from 100 kHz to 100 MHz. The thick gray line indicates the adiabatic limit calculated from the DC raw data using Eq.~1.
\textbf{d.} Same as \textbf{a} for frequencies ranging from 250 MHz to 6 GHz. The adiabatic limit (thick gray line) is shown for comparison.
\textbf{e.} Evolution of the relative amplitude change $\Delta I_{\rm{AB}}$ with frequency calculated using Eq.~(\ref{eq:deltaI}). The dashed line serves as a guide to the eye. Below 1 GHz, $\Delta I_{\rm{AB}}$ remains small, indicating adiabatic behavior. Above 1 GHz, $\Delta I_{\rm{AB}}$ deviates from the adiabatic limit, marking the transition to the non-adiabatic regime.}
\label{fig:fig3}
\end{figure*}

%
We now investigate the frequency response of our electronic MZI under sinusoidal drive at frequencies above 100 MHz as shown in Fig.\,\ref{fig:fig3}b. 
Contrary to the case at low frequency, 
as we increase the frequency, a distinct deviation from the adiabatic regime is observed, manifesting itself at frequencies
around 1~GHz. 
This is further supported by our Floquet simulations, which show a similar evolution towards the non-adiabatic regime at similar frequencies as the ones observed in the experiment (see Supplementary Note section 2.4).
These simulations are based on Floquet scattering theory \cite{Moskalets2011}, which describes electron transport under periodic voltage drives. Under a time-dependent voltage $V(t)$, electrons acquire a time-dependent phase $\phi(t) = \frac{e}{\hbar}\int V(t)dt$. The Fourier components of the phase factor $e^{-i\phi(t)}$ yield the photo-assisted probabilities $P_n$, where $n > 0$ ($n < 0$) corresponds to the absorption (emission) of $n$ photons. These probabilities determine the AC transport properties and are used to calculate the rectified current in the system. 

Our Floquet simulations capture well the features observed in both the adiabatic and non-adiabatic regimes, particularly the emergence of a maximum as a function of voltage bias and the deviations from the adiabatic regime at frequencies comparable to those observed experimentally.
However, the peak position of the coherent oscillations shifts in the opposite direction compared to the experimental observations (see Supplementary Note, Section 2.4). Differences are also observed at high bias voltages.
We attribute these discrepancies to the limitations of the Floquet scattering approach, which does not account for electron-electron interactions, hence decoherence. 
This is significant, as wavepackets generated by ultrashort voltage pulses are strongly affected by electron interactions, as demonstrated in \cite{Roussely2018}.
%
%
%
%

To further highlight the frequency-dependent evolution of the oscillation amplitude, we define $\Delta I_{\rm{AB}}(f)$ as the absolute difference between the maximum oscillation amplitude measured at a given drive frequency $f$ and the maximum oscillation amplitude in the adiabatic limit: 
\begin{equation}
\Delta I_{\rm{AB}}(f)= \\
\left| \max\left[ I_{\rm{AB}}(f \rightarrow 0, V_{\rm{ac}}) \right] - \max \left[ I_{\rm{AB}}(f, V_{\rm{ac}}) \right] \right|
\label{eq:deltaI}
\end{equation}
This quantity is plotted in Fig.\,\ref{fig:fig3}e.
At low frequencies, the quantum oscillations follow the adiabatic limit up to roughly 1~GHz. 
Beyond this frequency, the system presents deviations from the adiabatic limit and gradually
evolves towards the non-adiabatic regime. 

Notably, accessing the non-adiabatic regime has remained elusive until now, as it was expected to occur at frequencies well above 1~GHz \cite{Rossignol2018}. 
We demonstrate, that in our device, this regime is reached at surprisingly low frequencies due to the specific properties of the TCW's conduction modes. Using realistic electrostatic potential simulations, as pioneered in \cite{Chatzikyriakou2022} (Supplementary Note, section 2.4), we show that the modes near the Fermi energy exhibit strong energy dependence, leading to non-linear transport and, consequently, a rectified current. Owing to their low kinetic energy, these modes dominate the frequency response of our device, causing deviations from the adiabatic regime even at frequencies around 1 GHz.
%
%
%
%
%

\textbf{Electronic interference with on-demand single-electron wavepackets.}
We now demonstrate quantum interference beyond the adiabatic regime with ultrashort wavepackets having a temporal width as short as 30 ps and containing as few as one electron.
It is important to emphasize that, in our case, the generated wavepackets are plasmonic pulses influenced by electron interactions \cite{Roussely2018}.
To begin, we characterise these ultrashort plasmonic wavepackets on-chip using a pump-probe technique.
We apply a voltage pulse with a duration of 25 ps \cite{Aluffi2023} to the injection ohmic contact via the AC port of a high-bandwidth bias tee. 
By applying a second short voltage pulse to the quantum point contact, highlighted in white in Fig. 1, with a precisely controlled time delay, we can measure the time-resolved trace of the plasmonic wavepacket directly on chip. 
Time traces for various pulse amplitudes,
shown in Fig. 4b, 
reveal plasmonic pulses with a temporal duration of 30 ps.
For such short pulses, we observe quantum oscillations in the output currents of our Mach-Zehnder interferometer device, which are perfectly anti-phased. These interference patterns exhibit minimal background current and remain highly robust under applied bias voltages of up to several hundred microvolts.
%

\begin{figure*}[h!]
\centering
\includegraphics[scale=1]{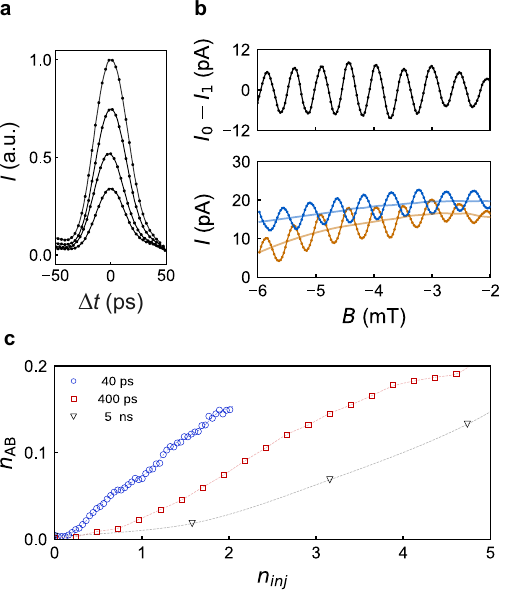}
\caption{
\textbf{Quantum interference with on-demand single-electron plasmonic pulses. }
\textbf{a.} Time-resolved measurements of a plasmonic pulse of width $\tau_{p}=30$~ps. The envelope is measured in a pump-and-probe experiment by varying the pulse amplitude.
\textbf{b.} Bottom panel shows raw data of Aharonov-Bohm oscillations for the shortest pulse for an amplitude of 100 $\mu$V. 
Top panel shows the total oscillating current $I_0$-$I_1$, where a smooth background has been subtracted.
\textbf{c.}  
Average number of transmitted charges $n_{AB}$ that contribute to the quantum coherence as a function of injected charges $n_{inj}$, with pulses of temporal widths $\tau_p$ varying from 40 ps to 5 ns. $n_{AB} = n_0 - n_1 $ corresponds to the total number of transmitted charges from the upper and lower electron waveguide that contribute to the quantum coherence. The dashed lines are guides to the eye.}

\label{fig:fig4}
\end{figure*}
To demonstrate that quantum interference can be observed with the on-demand injection of a single electron — a key requirement for realizing a flying electron qubit — we analyse our results in terms of the number of injected and transmitted electrons. 
We convert the amplitude of the coherent oscillations $I_{\rm{AB}}$, as defined above,
into the average number of interfering charges $n_{\rm{AB}}$ using the relation $I \,=\,enf$, where $e$ is the electron charge, $n$ is the average number of charges, and $f = 100$~MHz is the pulse repetition frequency. 
Similarly, we convert the pulse amplitude $V_{\rm{p}}$ into an effective number of injected electrons per pulse $n_{inj}$ (see Supplementary Note, Section 1.5). These results are presented in Fig.~\ref{fig:fig4}c for pulses of various temporal widths.
We observe that, with ultrashort voltage pulses, it is straightforward to reach a regime where a single electron traverses the interferometer. 
Remarkably, the contrast of the oscillating signal is significantly enhanced for shorter voltage pulses. 
This enhancement is primarily attributed to the high-energy components of short plasmonic pulses, 
which probes a higher energy range of the $I-V$ characteristic and is thus more sensitive to the nonlinearity.
A detailed understanding of the enhanced contrast of the coherent oscillations compared to the DC regime is currently lacking. 
Achieving this would require a microscopic theory that incorporates electron-electron interactions, which is computationally too costly at present.

Finally, let us comment on the relationship between the size of the plasmon pulse and the dimensions of the interferometer. 
As demonstrated above, the nonlinearity primarily stems from modes near the Fermi energy in the tunnel-coupled wire, which determine the effective propagation speed of the plasmon pulse. 
These modes exhibit the slowest Fermi velocity, $v_{TCW} \approx 3 \times 10^4$\,ms$^{-1}$ while the plasmon propagates at a speed of $\approx$ $10^6$\,ms$^{-1}$ in the interferometer arms. 
Assuming these velocities for the two different sections of the MZI, the resulting total propagation time is calculated to be 144 ps.
This timescale corresponds to half the period of a sine wave with a frequency of 3.5 GHz, matching well the frequency range where deviations from the adiabatic regime are observed. 
These insights suggest that the plasmon wavepacket is significantly smaller than the quantum device, supporting the consistency of our observations.




\section*{Conclusion:}

In conclusion, we have demonstrated electronic quantum interference in a 14-micron-long Mach-Zehnder interferometer using plasmonic pulses containing a single charge. 
By employing GHz sinusoidal excitation and voltage pulses with durations of several tens of picoseconds, we identified a new `non-adiabatic' regime. 
This achievement represents a significant milestone towards realizing flying electron qubits, where the pulse width must be shorter than the dimensions of the quantum device.

Beyond providing the first proof-of-principle demonstration of coherent control for ultrashort electron qubits in semiconducting systems, we expect that our detailed high frequency investigation will stimulate further theoretical and experimental research into the electron dynamics of these systems.
To complete the implementation of a fully-fledged flying electron qubit, the integration of single-shot detection is essential. A recent advancement has been achieved in this direction \cite{Thiney2022}.  
The next crucial milestone is to increase the number of flying qubits that can be accommodated within a single processing unit, enabling the implementation of multiple gate operations during their flight. This can be achieved by further reducing the temporal width  of the plasmonic pulses \cite{Whitney2024}, potentially reaching durations in the terahertz regime \cite{Giorgos2020}.

Furthermore, our demonstration of coherence in the non-adiabatic regime opens new possibilities for electron quantum optics experiments. 
This achievement paves the way for exploring dynamical interference control \cite{Gaury2014, Gaury2014_2} and provides new avenues for investigating single-electron coherence \cite{Haack2011} and multi-particle interference phenomena in electronic MZIs \cite{Hofer2014, Rossello2015, Kotilahti2021}. 
The coherent manipulation of ultrashort wavepackets in an electronic MZI also represents a crucial step towards coupling multiple interferometers to study entanglement and test Bell inequalities \cite{Bertoni2002, Vyshnevyy2013}. 
Moreover, these short wavepackets offer a pathway to high-fidelity quantum teleportation of single-electron states \cite{Olofsson2020}, marking a significant advance toward quantum information processing with flying electrons.

\newpage
\subsection*{Methods}
\label{sec:methods}


{\bf Sample fabrication.} 
The sample is patterned on a GaAs/AlGaAs heterostructure, forming a two-dimensional electron gas (2DEG) located 145~nm below the surface, with carrier density $n_e = 1.9 \times 10^{11} \,\text{cm}^{-2}$ and mobility $\mu_e = 1.8 \times 10^6 \,\text{cm}^2/\text{V}\cdot\text{s}$. The electronic Mach-Zehnder interferometer (MZI) geometry is defined by Ti/Au surface gates patterned using electron beam lithography. Electrical connections to the 2DEG are established through Ohmic contacts formed by the successive deposition of Ni(5~nm)/Ge(140~nm)/Au(280~nm)/Ni(100~nm)/Au(15~nm), followed by annealing under a continuous flow of forming gas (5\% H$_2$ in Ar) at 370$^\circ$C for 2 minutes and 430$^\circ$C for 1 minute. The injection Ohmic contact area measures 10×10~µm$^2$.
%

{\bf High frequency signal injection.}
The high frequency signal (sinusoidal or pulses) was
injected into the MZI through  coaxial line with a bandwidth of 40 GHz.
It was modulated at 170 Hz and  injected into a high bandwidth (40 GHz) bias-tee to control independently the AC and DC components of the signal. 
The ultrashort voltage pulses used in this experiment were generated using a homemade voltage pulse generator based on frequency comb synthesis \cite{Aluffi2023}, in conjunction with an arbitrary waveform generator with a sampling rate of 24 GS/s. 
The time-resolved measurement of the 30 ps pulse confirms the precise injection of ultrashort voltage pulses into the MZI without significant distortion.

{\bf Electrostatic and quantum transport simulations.}
Semi-quantitative quantum transport simulations are performed in two steps. 
First, the electrostatic potential in the silicon doped heterostructure is calculated by solving the Poisson equation with the commercial solver nextnano \cite{nextnano}.
We follow the approach adopted in \cite{Chatzikyriakou2022}.
To implement the device geometry, we position metallic gold gates on the surface of GaAs crystal, taking into account the Schottky barrier. 
Surface charges are added to take into account Fermi-level-pinning. 
The dopant and the surface charge densities are calibrated in such a way that they reproduce an experimental pinch-off measurement between two metallic surface gates. 
In a second step, a 2D slice of the electrostatic potential at the 2DEG height is extracted and used to compute the DC current. 
The Landauer-B\"uttiker formalism \cite{Datta}
is employed, and the calculations are performed using the open-source software Kwant \cite{Groth2014}. 
The lattice constant is set to $a = 5$\,nm, and the magnetic field is incorporated using the standard Peierl's substitution. 
Additionally, the current under periodic drive is calculated from the DC current using Floquet scattering theory, as described in \cite{Rossignol2018}.

\subsection*{Data availability}
The datasets used in this work will be made available online from the Zenodo repository.

\subsection*{Acknowledgements}
The authors acknowledge fruitful discussions with P. Degiovanni during the preparation of this manuscript.
This project has received funding from the European Union H2020 research and innovation program under grant agreement No. 862683, “UltraFastNano”. 
C.B. acknowledges funding from the French Agence Nationale de la Recherche (ANR), project ANR QCONTROL ANR-18-JSTQ-0001.
C.B., H.S., X.W. and J.S. acknowledge funding from the Agence Nationale de la Recherche under the France 2030 programme, reference ANR-22-PETQ-0012. 
J.W. acknowledges the European Union H2020 research and innovation program under the Marie Sklodowska-Curie grant agreement No. 754303. 
M.A. acknowledges the MSCA co-fund QuanG Grant No. 101081458, funded by the European Union and the program QuanTEdu-France n° ANR-22-CMAS-0001 France 2030.
L.M. acknowledges the program QuanTEdu-France n° ANR-22-CMAS-0001 France 2030.
T.V. acknowledges funding from the French Laboratory of Excellence project "LANEF" (ANR-10-LABX-0051).
M.Y. acknowledges CREST-JST (grant number JPMJCR1675) and JST Moonshot (grant numbers JPMJMS226B-4). 
M.Y., D.P., and S.T. acknowledge Japan Society for the Promotion of Science, Grant-in-Aid for Scientific Research S (grant number 24H00047).
S.T. acknowledge CREST-JST (grant number JPMJCR1876) and JST Moonshot (grant numbers JPMJMS226B).
G.G. acknowledges EPSRC ``QUANTERAN" (grant number EP/X013456/1) and Royal Society of Edinburgh ``TEQNO"  (grant number 3946).
A.D.W. and A.L. thank the DFG via ML4Q EXC 2004/1 - 390534769, the BMBF-QR.X Project 16KISQ009 and the DFH/UFA Project CDFA-05-06.
The present work has been done in the framework of the International Research Project “Flying Electron Qubits” – “IRP FLEQ”- CNRS – Riken – AIST – Osaka University.

Views and opinions expressed are those of the author(s) only and do not necessarily reﬂect those of the European Union or the granting authority. Neither the European Union nor the granting authority can be held responsible for them.

\subsection*{Author contributions}
S.O. fabricated the sample with help from G.G. and performed the experiment with support from M.A., T.V., H.E. and J.W. and technical assistance from C.G., J.S. and M.U. 
\hspace{1mm}
L.M. and T.K. performed the numerical simulations with help from X.W.\hspace{2mm}
M.Y., D.P., S.T, and N-H. K. helped in the interpretation of the experimental data. 
G.G. has participated in the early stage of the project. 
A.L. and A.D.W. provided the high-quality GaAs/GaAlAs heterostructure. 
S.O., L.M. and T.K. wrote the manuscript with feedback from all authors. H.S and C.B. supervised the experimental work. C.B. has initiated the project.

\printbibliography[title={References}]
\end{refsection}

\pagebreak

\;\vfill
\begin{center}
    \textsf{\textbf{\Huge Supplementary Notes\\\vspace{2mm}}}
\end{center}
\vspace{-2mm}
\def\einr{2mm}
\def\spazi{-1.7mm}

\vfill 
\tableofcontents 
\thispagestyle{empty}
\vfill
\pagebreak

\section{Experimental techniques and sample characterisation}

\subsection{Experimental setup }
\begin{refsection}
The high-frequency AC signal is injected into a dedicated Ohmic contact of the sample using a high-frequency transmission line and a high bandwidth bias-tee (SHF BT 45 A) to
independently control the AC and DC component of the signal.
The output currents are measured via the voltage drop across a 10~k$\Omega$ resistor to ground, located on the chip carrier at a temperature of 30~mK.
The AC signal is modulated at a frequency of 170 Hz to perform lock-in technique measurements.

In our measurement setup, the sample's output current can be modeled as a current source in parallel with the sample's resistance. While a transimpedance amplifier would typically be used for precise current measurements, these amplifiers present several challenges in our experimental conditions. They tend to inject noise at their input and are not rated for operation at cryogenic temperatures.
To overcome these limitations, we implemented a simpler solution using 10~k$\Omega$ resistors as cold grounds. These resistors are soldered directly onto the chip carrier at the lowest temperature stage of the dilution refrigerator, providing minimal thermal noise. However, this approach creates a voltage divider between the sample resistance and the 10~k$\Omega$ resistor, resulting in a reduction of the effective bias applied to the sample.
This bias reduction has been taken into account by considering the voltage division ratio. 
This correction factor has been systematically applied to all bias-dependent measurements presented in this work.

\subsection{Tuning of anti-phase Aharonov-Bohm oscillations}
\begin{figure*}[h!]
    \centering
    \includegraphics[scale=1]{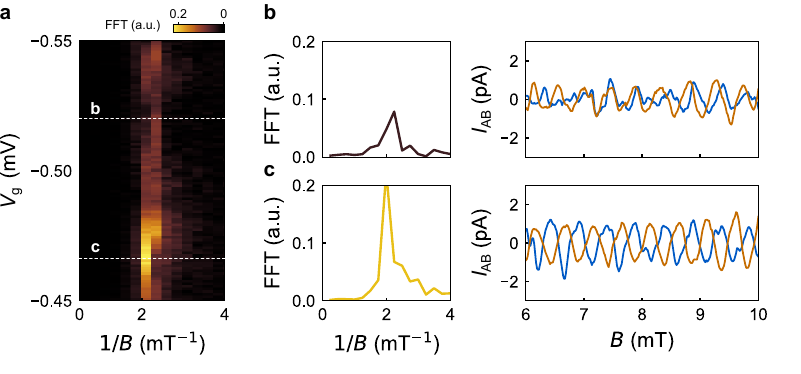}
    \caption*{\textbf{Figure S1: Optimisation of anti-phase AB oscillations.}
    \textbf{a.} 
    FFT intensity plot of the difference between the oscillating components  of the currents \(I_{0}\) and \(I_{1}\) as a function of the gate voltage \(V_{\text{g}}\). 
    \textbf{b.}
    FFT of the current difference \(I_{0} - I_{1}\) (left) and the oscillating components of \(I_{0}\) and \(I_{1}\) as a function of the magnetic field \(B\), corresponding to configuration (1) in \textbf{a}, where low FFT intensity indicates in-phase oscillations that are not well-tuned. \textbf{c.} Same as \textbf{b} but for configuration (2), where high FFT intensity indicates well-tuned anti-phase oscillations.}
\end{figure*}

A notable characteristic of our Mach-Zehnder interferometer is its operation at low magnetic fields (few mT), devoid of quantum Hall effect and chirality-related phenomena. 
The optimal operation of the interferometer requires precise tuning of the electrostatic gates to achieve the two-path regime, where contributions from multiple-path trajectories (such as paths encircling the AB ring) are suppressed
\cite{Yamamoto2012,Takada2015}.
The gates are usually operated between the 2D pinch-off where complete depletion occurs under individual gate, and the pinch-off between two  adjacent gates. This tuning process involves sweeping the magnetic field $B$ while monitoring the AB oscillations in both output currents $I_0$ and $I_1$. By adjusting the gate voltages within this working range, we maximize the FFT amplitude of the current difference $I_0 - I_1$, ensuring well-defined anti-phase oscillations \cite{Yamamoto2012,Takada2015},as illustrated in Fig. S1c. This optimization procedure is performed independently for each gate.

From the magnetic field periodicity of the current oscillations ($\Delta B = 0.5$~mT), 
we can extract the effective area $S$ using the relation $S = h/e\Delta B$, yielding $S = 8.2$~$\mu$m$^2$. This experimental value can be compared with the geometric constraints of our device. Given the AB ring length of 10~$\mu$m and accounting for the depletion length of 50~nm from the surface gates, the path width varies from 300~nm (inner trajectory) to 1000~nm (outer trajectory). These dimensions correspond to possible enclosed areas ranging from 3~$\mu$m$^2$ to 10~$\mu$m$^2$. The experimentally extracted area falls within this range and is closer to the upper bound, consistent with the ballistic nature of transport where electrons predominantly follow outer trajectories.

\subsection{Time-resolved characterisation of the plasmonic pulse}
\begin{figure*}[h!]
\centering
\includegraphics[scale=1]{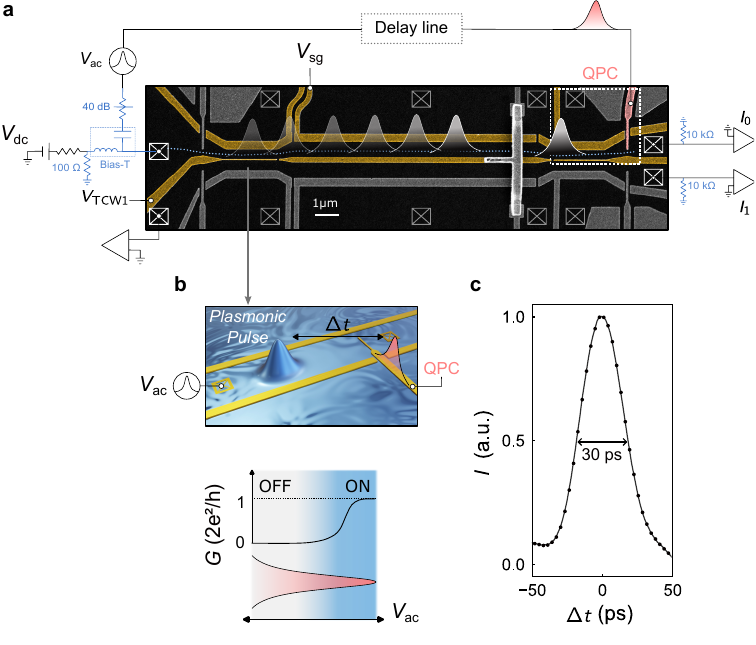}
\caption*{\textbf{Figure~S2: Time resolved characterisation of the plasmonic pulse as a function of the pulse amplitude}.
\textbf{a.} SEM image of the MZI device and schematic of the experimental setup used for time-resolved measurements at 30~mK.
\textbf{b.} (Upper panel) Artistic representation of the propagating plasmonic pulse and pump-probe measurement scheme. (Lower panel) Operating principle of the QPC as a fast switch: the QPC, initially in pinch-off regime (zero conductance) due to applied negative DC bias, is momentarily opened by an ultrashort voltage pulse, allowing current transmission.
\textbf{c.} Time-resolved detection of a 30~ps plasmonic pulse.}  
\end{figure*}

To characterise the injected ultrashort plasmonic pulse in the device, we perform time-resolved measurements pulses injected at 3~GHz in the upper-left Ohmic contact, as shown in Fig.~S2a. We bias negatively all electrostatic gates that define the upper path of the MZI, including the middle gates (central island, $V_{\rm{TCW1}}$, and $V_{\rm{TCW2}}$), guiding the plasmonic pulse along the upper arm of the interferometer towards the upper-right Ohmic contact.

The measurement setup uses a power divider to split the output signal of our homemade voltage pulse generator \cite{Aluffi2023}. One part is sent to the AC injection Ohmic contact on the sample, while the other passes through a computer-controlled mechanical delay line before reaching the QPC. 
Both the Ohmic contact line and the QPC line are equipped with 40 GHz bandwidth bias-tees (SHF BT 45 A). The QPC is initially biased in the pinch-off regime, acting as a fast switch as depicted in Fig.~S2b. An ultrashort voltage pulse with positive amplitude opens the QPC momentarily, with a switching time significantly shorter than the electron wavepacket duration, enabling accurate temporal reconstruction of the signal.
By sweeping the time delay $\Delta t$ between the injected plasmonic pulse and the pulse sent to the QPC, we measure the time-resolved current using lock-in detection, as shown in Fig.~S2c. The measured pulses exhibit minimal distortion, demonstrating high-quality transmission and effective injection of ultrashort plasmonic pulses in our quantum device.

\subsection{Temperature dependence of coherent oscillations in the DC regime}
\begin{figure*}[h!]
\centering
\includegraphics[scale=1]{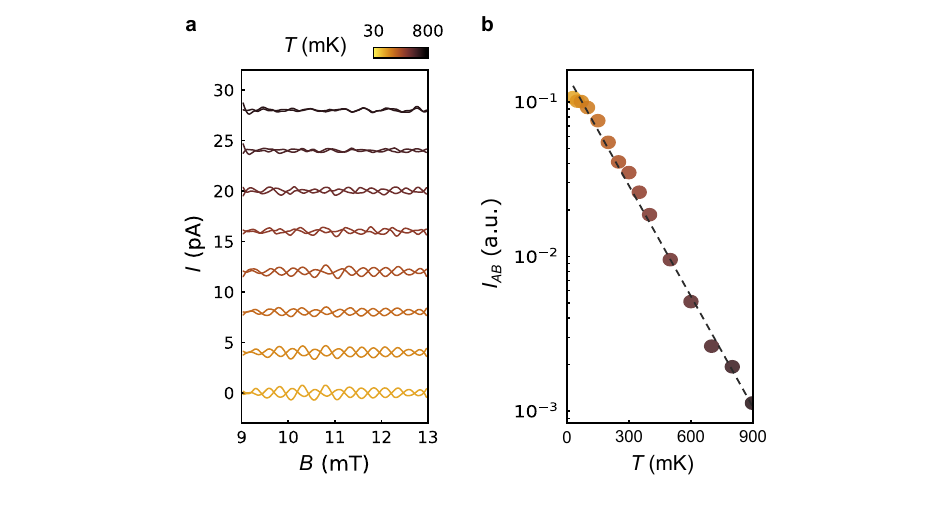}
\caption*{\textbf{Figure~S3: Temperature dependence of AB oscillations in the DC regime.} \textbf{a.} Current $I_0$ and $I_1$ as a function of magnetic field $B$ for different temperatures $T$. A vertical offset has been added for clarity.
\textbf{b.} Amplitude of the FFT of the current difference \(I_{0} - I_{1}\) as a function of temperature.}  
\end{figure*}
The coherence length $l_{\phi}$ can be estimated by measuring the temperature dependence of the AB-oscillation amplitude \cite{Kobayashi2002,Yamamoto2012}. 
Fig.~S3a shows the temperature dependence of the Aharonov-Bohm oscillations of our device in the DC regime. 
Following the procedure of Yamamoto \textit{et al.} \cite{Yamamoto2012}, we evaluate the coherence length $l_{\phi}$ to 80 µm for our 14-µm-long Mach-Zehnder interferometer at 30 mK.

\subsection{Estimation of the number of electrons per pulse}
To interpret the data in terms of the number of electrons, we first estimate the charge contribution from the positive part of the square pulse. In an ideal linear system, the positive and negative components of the pulse would result in zero net charge transfer. However, quantum rectification in our system generates a non-zero average current, justifying a focus on the positive portion of the pulse to estimate the number of injected electrons.
The square pulse signal (with no DC component, \(\langle V(t) \rangle = 0\)) is defined as:
\begin{equation}
V(t) = 
\begin{cases} 
    \left(1 - \frac{\tau}{T}\right) V_{\text{p}}, & 0 \leq t \leq \tau, \\ 
    -\frac{\tau}{T} V_{\text{p}}, & \tau < t \leq T,
\end{cases}
\end{equation}
where \(\tau\) is the pulse width, \(T\) is the period, and \(V_{\text{p}}\) is the amplitude of the pulse. 

For simplicity, the number of electrons injected by the positive part of the pulse is computed assuming a single conductance channel, with the current expressed as \(I(t) = \frac{e^2}{h} V(t)\). Although this assumption does not strictly fulfilled for our system, as it involves multiple transmitting channels, it is justified because the dominant contribution to transport comes from few modes near the Fermi level, which exhibit strong energy dependence (see Figs. S8 and S10). 

The average number of injected charges is given by:
\begin{equation}
n_{inj} = \int I(t) \, dt = \frac{e}{h} \int_0^\tau \left(1 - \frac{\tau}{T}\right) V_{\text{p}} \, dt = \frac{e}{h} (1 - \alpha) V_{\text{p}} \tau,
\end{equation}
where \(\alpha = \frac{\tau}{T}\).
The average measured current \(I\) can then be related to the average number of injected charges \(n_{inj}\) and the repetition frequency \(f = 1/T\) (100 MHz) through the relation:
$
I = e\,n_{inj}\,f.
$

\subsection{Measurement of the nonlinearity of the tunnel-coupled wire}

\begin{figure*}[h!]
\centering
\includegraphics[scale=0.8]{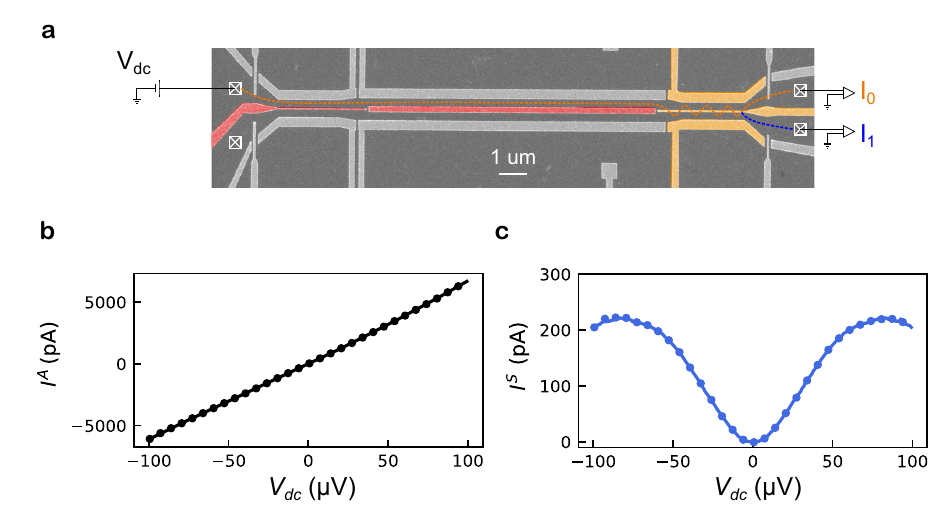}
\caption*{\textbf{Figure~S4: Nonlinearity of the tunnel-coupled wire.} 
\textbf{a.}
SEM picture of the new MZI device. The gates in red are strongly polarized to separate the upper path from the lower path
\textbf{b.} 
Antisymmetric component of the current $I_0 - I_1$ as a function of bias voltage $V_{dc}$. 
\textbf{c.} 
Symmetric component of the current $I_0 - I_1$ as a function of bias voltage $V_{dc}$.}  
\end{figure*}
In the manuscript, we attribute the nonlinearity to the tunnel-coupled wire.
In addition to being supported by our numerical simulations, this claim is further substantiated by additional data from another electronic Mach-Zehnder interferometer device, as shown in Fig.~S4a, where we specifically investigate the nonlinearity of the TCW alone.
While the geometry of the TCW on this sample remains identical to that described in the main paper, its length has been extended from 2 $\mu m$ to 3 $\mu m$.
We apply sufficiently strong voltages to the gates highlighted in red, effectively separating the upper and lower path and we inject the current in the upper part of the TCW. 
We measure the output currents $I_0$ and $I_1$ while sweeping the bias voltage $V_{dc}$ applied to the upper channel.
Similarly to what has been done in the main manuscript, we decompose the output current into its symmetric (Fig.~S4c), $I^{S} (V,B) =
 \big (I(V,B) + I(-V,B) \big )$\,/\,2,
and antisymmetric (Fig.~S4b), $I^{AS} (V,B)  =
 \big ( I(V,B) - I(-V,B) \big )$\,/\,2,
components. 
We observe a very similar behaviour as in the MZI, confirming that the TCW is at the origin of the observed the nonlinearity.



\section{Numerical modelling of the device}
The following section summarises the numerical transport simulations of a realistic device model, which were performed to help interpreting the experimental findings.
An actual simulation consists of two steps:
First, the electrostatic potential of the heterostructure is obtained for a specific configuration of gate voltages by solving the Poisson equation.    
Second, transport calculations of the 2DEG are performed using a tight-binding ansatz in combination with the previously calculated electrostatic potential. The electrical current through the interferometer is then easily obtained within the Landauer-B\"uttiker formalism \cite{Datta}. 
Both steps are described in more detail in the following.

\subsection{Electrostatic potential simulations}
\begin{figure*}[h!]
\centering
\includegraphics[scale=1]{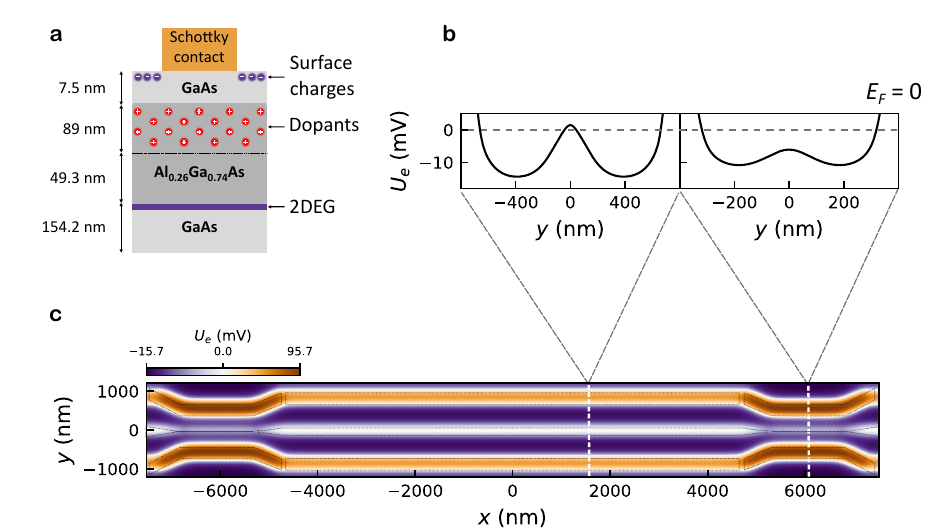}
\caption*{\textbf{Figure~S5: Electrostatic simulations of the MZI.} \textbf{a.} Schematic representation of the layered device along the height in $z$-direction. The gates are modelled by Schottky contacts and surface charges were added in the ungated region. The Fermi energy at the 2DEG depth is set to $E_F = 0$ eV. \textbf{b.}  Computed $U_e$ along a 1D cut through the TCW and the AB region. \textbf{c.} Top view of $U_e$ in the full device region. The positions of the gates at the surface are indicated by the dotted line polygons.
} 
\label{fig:supp_fig1}
\end{figure*}
The Poisson equation which relates the electrostatic potential $U(\Vec{r})$ at position $\vec{r} = (x, y, z)^T$ to the charge density $\rho(\Vec{r})$
is
\begin{equation}
    \Vec{\nabla}\cdot\biggl[ \epsilon(\Vec{r})\Vec{\nabla}U(\Vec{r}) \biggr]
    = \rho(\Vec{r}).
    \label{eq:poisson}
\end{equation}
In above formula, $\epsilon$ is the dielectric constant which has an explicit spatial dependence, as the device is a layered AlGaAs/GaAs heterostructure, as shown in Fig.~S5a.
We employ an ansatz similar to the one used in reference \cite{Chatzikyriakou2022}, which has been shown to quantitatively reproduce experimental pinch-off data.
For this, the charge density $\rho$ is modelled by three different contributions $\rho/e = N_{2DEG} - N_d + N_s$, where
$N_{2DEG}$ is the electron density inside the 2DEG layer, $N_d$ the density of dopant charges and $N_s$ the surface charge density in the ungated region.
Using Thomas-Fermi approximation, one can relate the electron density inside the 2DEG to the electrostatic potential $U_e(r) \equiv U(x, y, z_{2DEG})$ in the 2DEG plane at height $z_{2DEG}$ and $r = (x, y)^T$ is a vector in the plane of the 2DEG. 
At zero temperature
\begin{equation}
    N_{2DEG}(r) = \frac{(2m^*)^{3/2}}{3\pi^2\hbar^3}( E_F + eU_e(r) - E_c)^{3/2} \,\,\, \text{for} \ E_F +eU_e(r) > E_c \  \text{and 0 otherwise},
    \label{eq:thomas_fermi}
\end{equation}
where $E_c$ is the bottom of the conduction band, $m^*$ is the effective electron mass which is set to $0.067 m_e$ for GaAs and $E_F$ the Fermi energy which is zero in our case.
The calibration procedure for the two parameters $N_s$ and $N_d$ is explained in the next section. Electrostatic gates are taken into account via Neumann boundary conditions, while the interface between gates and the semiconductor are represented by Schottky contacts. 
The numerical solution of above self-consistent equations (\ref{eq:poisson}) and (\ref{eq:thomas_fermi}) are performed with the nextnano++ software \cite{nextnano}.

For the subsequent transport simulation, we are especially interested in electrostatic potential $U_e$ at the 2DEG layer.
Fig.~S5c shows this potential after calibration of the model in the full interferometer device while Fig.~S5b shows a cut along $y$-direction. The areas of the 2DEG where the potential energy is above the Fermi energy $E_F = 0$ eV are depleted and form the interferometer geometry with a quasi-1D waveguide.

\subsubsection{Calibration procedure}

\begin{figure*}[t!]
\centering
\includegraphics[scale=1]{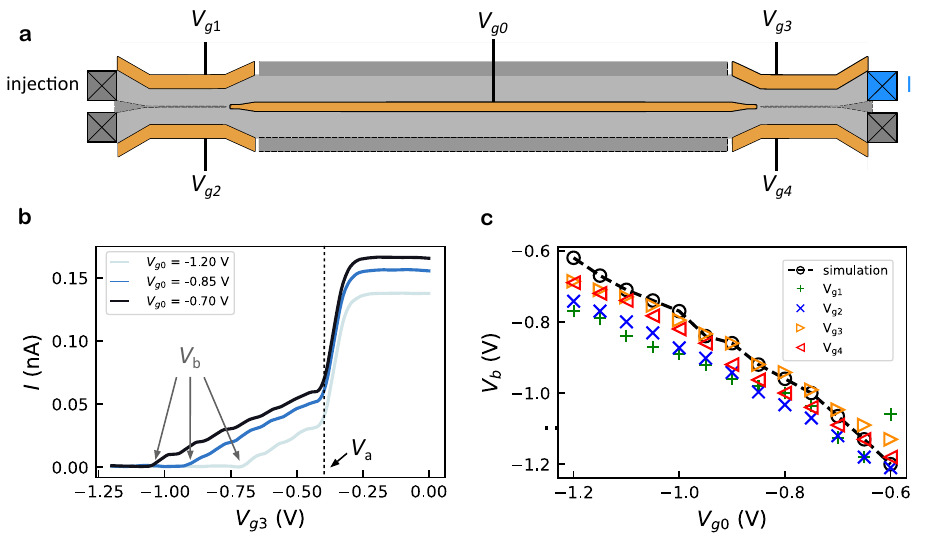}

\caption*{\textbf{Figure~S6: Calibration of the electrostatic model against experimental pinch-off data.}
\textbf{a.} Schematic device representation. 
The gates highlighted in orange are selected to measure pinch-off curves. 
The current $I$ is measured at the blue ohmic contact.
\textbf{b.} Experimentally measured current $I$ as a function the gate voltage $V_{g0}$ for three different values of $V_{g3}$. 
The characteristic value of $V_a$, where the slope changes, and of $V_b$, where the current drops to zero (pinch-off voltage), are extracted and used for calibration and verification of the electrostatic model. 
\textbf{c.} Comparison between the experimental and simulated pinch-off voltages $V_b$ as a function of the gate potential $V_{gi}$ for $i = 1,2,3 \ \rm{and} \ 4$ for the calibrated model.
} 
\label{fig:supp_fig6}
\end{figure*}

Our electrostatic model contains two \textit{a priori} unknown parameters, the surface charge density $N_s$ and the dopant density $N_d$.
 We apply the same recipe as in \cite{Chatzikyriakou2022} and determine both values by fitting experimental pinch-off data. 
 This procedure has proven to be successful to obtain a quantitatively precise model without additional fitting parameters.

 The calibration procedure is iterative.
 First, a pair of neighbouring gates is chosen, such that both lie above and below the (upper or respectively lower) waveguide. For the subsequent discussion, we choose gate $g0$ with one of the gates $V_{gi}$ for $i = \{1,2,3,4\}$ as highlighted in orange in Fig.~S6a. 
 Experimental measurements of the current $I$ through the interferometer are then performed at the blue Ohmic contact as a function of the two gate potentials $V_{g0}$ and $V_{gi}$.
 The current $I$ as a function of $V_{g3}$ for three different values of $V_{g0}$ is shown in Fig.~S6b. 
 The shape of this curve is qualitatively similar for other gate combinations and has been analysed in detail in Ref.\ \cite{Chatzikyriakou2022}. 
 When the potential $V_{g3}$ is decreased from roughly $-0.3$ to $-0.4$\,V, the 2DEG below the gate is depleted which manifests in the steep decrease of $I$. 
 At voltage $V_a$ the 2DEG underneath the gate $V_{g3}$ is fully depleted. Decreasing further the potential at $V_{g3}$, the 2DEG in between the two orange gates becomes more and more depleted, until the current $I$ eventually drops to zero at the so-called pinch-off voltage $V_b$.
 
 The experimentally obtained value of $V_a$ is used in the following to determine $N_d$. For this, Eq.~\eqref{eq:thomas_fermi} is solved for the actual device model. 
 For a given value of $N_d$, the value of $N_s$ can be uniquely determined by requiring that the 2DEG density in the gated and the ungated region in similar. 
 These simulations have been performed on a simplified model containing just a single gate, which we choose in practice to be $200$\,nm wide. After this step, the value of $N_d$ can be uniquely determined by requiring that the 2DEG is fully depleted underneath the gate at voltage $V_a$.

The fitting procedure described above is repeated for all relevant gate pairs and leads to slightly varying values of $N_s$ and $N_d$. Due to that, the resulting 2DEG density at zero gate voltages is locally different.
We find that the 2DEG density varies around 10\% on a micrometer length scale, similar to what was observed in  Ref.\ \cite{Chatzikyriakou2022} and has been interpreted as charge disorder.
In the following simulations, disorder is not taken into account for simplicity 

 In a subsequent verification step, we apply the calibrated model to estimate the pinch-off voltages
 $V_b$, where the 2DEG is fully depleted between the gates $g0$ and $gi$.
 Fig.~S6c compares pinch-off voltages obtained from experimental measurements and numerical simulations. 
 We find an accuracy of $V_b$ below 0.1\,V for $g3$ and $g4$, and of around 0.15\,V for gates $g1$ and $g2$.

\subsection{Quantum transport simulation}

Numerical transport simulations were performed by using a microscopic tight-binding model on a square lattice with lattice constant $a$ similar as in Ref.\ \cite{Bautze}.
The Hamiltonian is
\begin{equation}  
H = \frac{1}{2m}[i\hbar\Vec{\nabla} - e \vec{A}(x,y)]^2 - e U_e(x,y)
\end{equation}
where $\vec{A}(x,y)$ is the magnetic vector potential taking into account the magnetic field $B(x,y)$ and  electrostatic potential $U_e(x,y)$ in the 2DEG calculated before.
Using standard Peierls substitution to account for the magnetic field, the nearest-neighbour coupling between two sites $i$ and $j$ with strength $\gamma$ is modified to $\gamma e^{i 2 \pi \phi / \phi_0}$, where $\gamma = \hbar^2 / (2 m a^2)$, $\phi_0 = h / e$ and 
$\phi = (y_j-y_i) (x_j+x_i) B$. In practice, we take $a = 5$\,nm and compute the conductance $G$ with the help of the open-source software Kwant \cite{Groth2014}.

The transport calculation of the full system can be accelerated by two tricks.
The first one consists in subdividing the full system into four sub-block as shown in
Figure S7.
If not yet present, each sub-block is extended in $x$-direction by semi-infinite leads. Due to the sub-block segmentation, the scattering matrix $S$ acquires a block structure. This has the advantage that only individual sub-blocks are affected by local gate voltage changes, preventing the expensive recalculation of the full scattering matrix.
To reconstruct the matrix of the full system, we first write a general scattering matrix with two leads in the form: 

\begin{equation}
    S_A = 
\begin{pmatrix}
r_{A} & t'_{A}  \\
t_{A} & r'_{A}  \\
\end{pmatrix}
\end{equation}

where $r_{A}$ ($r'_{A}$) is the reflection matrix corresponding to lead 1 (lead 2), and $t_{A}$ ($t'_{A}$) is the transmission matrix from lead 1 to lead 2 (from lead 2 to lead 1). 
The scattering matrix $S_{A+B}$ of two systems A and B in series is given by the Redheffer star product \cite{Raymond1959} as:

\begin{equation}
      S_{A+B} = \begin{pmatrix}
r_{A} + t'_A r_B\frac{1}{1-r'_A r_b} t_A & t'_{A} \frac{1}{1-r_b r'_A}t'_B  \\
t_{B} \frac{1}{1-r'_A r_B} t_A & r'_{B} + t_B r'_A \frac{1}{1-r_b r'_A} t'_B \\
\end{pmatrix}.
\end{equation} 

\begin{figure*}[h!]
\centering
\includegraphics[scale=1]{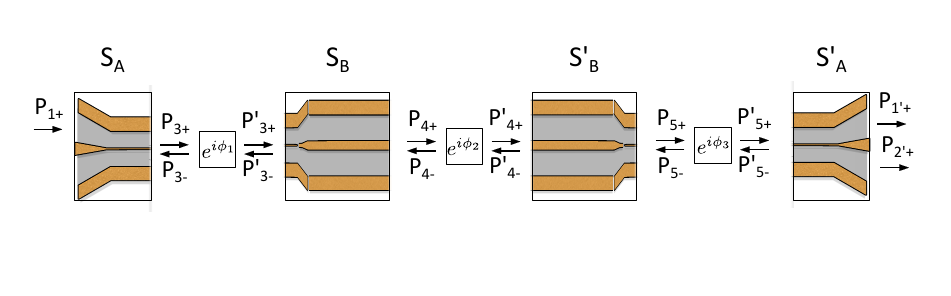}
\caption*{\textbf{Figure~S7: Schematic of the segmentation procedure.} To simplify the calculation of the scattering matrix for the MZI device, it is divided into 4 smaller sub-blocks for which the scattering matrices can be calculated independently. 
The AB phase is taken into account by adding a phase by hand in-between sub-blocks.
 }.
\label{fig:supp_fig7}
\end{figure*}

A second trick to speed up the calculation is to realise that the net-effect of the magnetic field is to introduce a phase difference between the modes propagating in the upper and in the lower path of the AB ring in the central block. 
Instead of performing a separate simulation for each individual $B$ value, it is sufficient to include the phase difference via $\phi_2$ in to the lower path only, see Figure S6. The phases $\phi_1$ and $\phi_3$ amount to change the interference due to the TCW, which have a dependence as $k l$, where $k$ is the longitudinal wavevector and $l$ the length of the coupling region.

Finally, the current from lead $j$ to lead $i$ is obtained from standard Landauer-Büttiker formula $I_{ij} = \int_0^{eV_{dc}}G_{ij}(E)dE$ where $V_{dc}$ is the voltage difference between both leads, $G_{ij} = |t_{ij}|^2$ is the conductance and we have assumed zero temperature.

\subsection{Non-linear behaviour of the beamsplitter}
Pure AC input signals to the interferometer device require rectification to result in a non-zero DC output current.
This is possible when the transmission is non-linearly dependent on energy. From our numerical scattering matrix simulations we find that the TCW in particular has a larger energy-dependent transmission. 
Fig.~S8a shows the lower current $I_1$ through the beamsplitter device and Fig.~S8b the corresponding transmission rates for three different tunnel barrier potentials $V_{\text{TCW}}$.
The transmission shows anti-phase oscillations between the upper and the lower path as a function of energy, resulting from interferences between propagating modes in the TCW \cite{Bautze}.
Writing the most dominant contribution in terms of a symmetric (S) and an antisymmetric (A) mode, the corresponding phases are $\phi_{S,A} = k_{S,A}L = \frac{1}{\hbar}\sqrt{2m_*(E-E_{S,A})}L$, where $k_{S,A}$ is the wavevector of the mode, $E_{S,A}$ is the confinement energy of the mode, $L$ is the length of the TCW and $E$ is the injection energy.
Let us note that non-linear behaviour in our model arises only from the scattering ansatz, without further more complicated mechanisms such as interactions. Previous studies where AB oscillations in the non-linear regime include electron-electron interactions have been reported in \cite{Angers2007,Leturcq2006}.
\begin{figure*}[h!]
\centering
\includegraphics[scale=1]{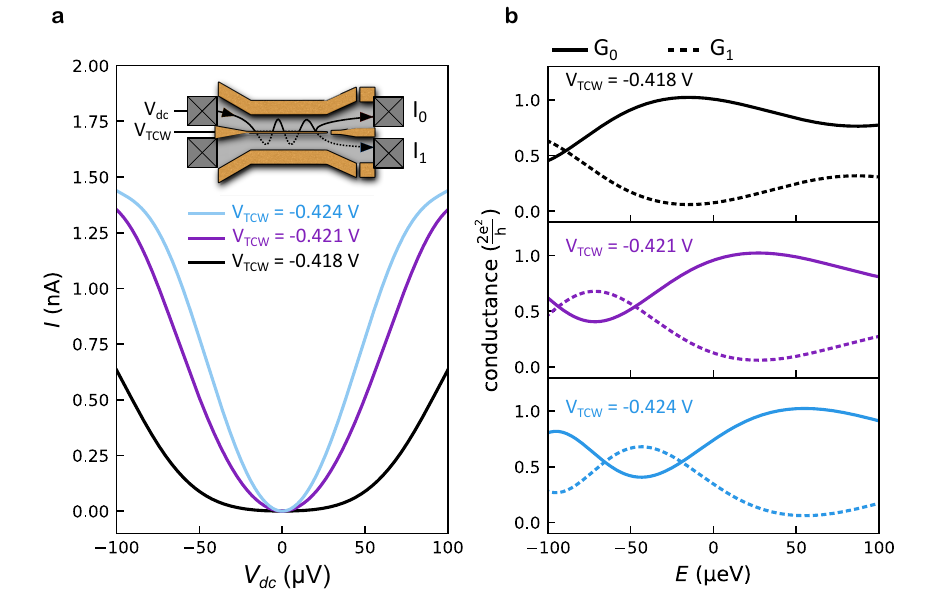}
\caption*{\textbf{Figure~S8: DC analysis of the TCW.} \textbf{a.} Symmetric part of the current $I_1$ from the upper-left to the lower-right lead as a function of $V_{\rm{dc}}$ for three values of $V_{\rm{TCW}}$. The inset shows a schematic representation of the device region. \textbf{b.} Conductance $G_{0/1}$ (label 0 refers to the upper-left to upper-right and label 1 to upper-left to lower-right conductance) as a function of the injection energy $E$ for three different values of $V_{\rm{TCW}}$. The non-oscillating component has been subtracted.} 
\label{fig:supp_simu_fig3}
\end{figure*}

 \subsection{Numerical simulation of sinusodial drives using Floquet scattering approach}

We use a Floquet scattering approach similar to \cite{Rossignol2018} to calculate the rectified current when driving the MZI with a sinusoidal voltage.
Injecting a potential $V(t) = V_{ac} \sin(\omega t)$ in the left lead, where $V_{ac}$ is the amplitude and $\omega$ the drive frequency, the incoming modes get multiplied by a factor of $e^{-i\phi(t)}$ due to the additional time-dependent phase $\phi(t) = \int_0^t \frac{eV(t')}{ \hbar }dt'$.
The phase factor is further decomposed as $e^{-i \phi(t)} = \sum_n P_n e^{-i \omega n t}$, where $P_n = J_n(\frac{eV_{ac}}{\hbar\omega})$ and $J_n$ are the Bessel functions of the first kind. 
Floquet scattering theory amounts to express the average current $\langle I(V) \rangle = \frac{\omega}{2 \pi} \int_0^{2 \pi / \omega} I(t)$ in terms of the DC conductances. At $T = 0$\,K, one finds
\begin{equation}  
\langle I(V) \rangle = \frac{e}{\hbar} \sum_n |P_n|^2 \int_0^{n\hbar\omega} G(E)dE.
\label{eq:floquet}
\end{equation}
For actual computations, above sum is truncated at a maximal $n$. We use the criterion that difference is below $10^{-3}$, when calculating the sum up to $n$ and $2 n$ elements. In practice, values up to $n = 10^4$ are needed.
The numerical integral is computed from a linear spline interpolation. 
For the conductance, the previously simulated DC values for $G_{0/1}(E)$ are used.
The amplitude of the AB oscillation is obtained by performing a Fourier transform for each $I$ vs.\ $B$ curve and taking the peak value of this curve. We show the result as a function of the driving strength $V_{ac}$ for different driving frequencies $\omega$ in Figure S9. 
The adiabatic limit is computed from the definition Eq.\ (2) in main text direct numerical integration over one time period. 
One finds that up to around 1\,GHz, the rectified current matches precisely the adiabatic limit, as shown in Fig.~S9a. 
Note that the simulated curves reproduce the main features of the experimental ones shown in Fig.~3, with 
a similar behaviour of $I_{AB}$ with $\omega$. For low drive frequencies up to 1 GHz, all curves show a pronounced maximum, which starts to deviate from the adiabatic limit as shown in Fig.~S9b and which is also found experimentally.
\begin{figure*}[h!]
\centering
\includegraphics[scale=0.75]{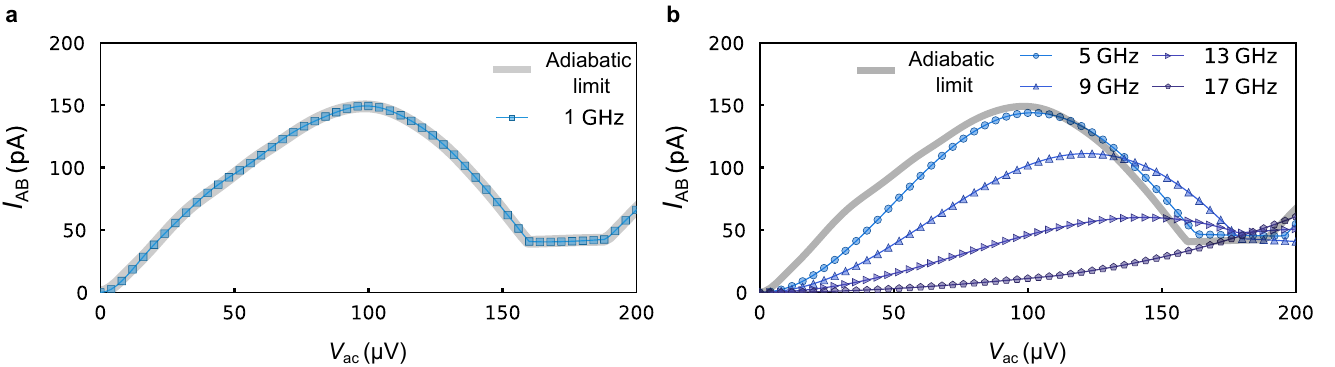}
\caption*{\textbf{Figure~S9: Simulation of the frequency response of the MZI under sinusoidal drive.} Amplitude of the coherent AB oscillations as a function of the amplitude of the sinusoidal drive $V_{\rm{ac}}$, obtained using the Floquet scattering approach in Eq.\ (\ref{eq:floquet}). 
\textbf{a.} Result for drive frequencies ranging from 100 kHz to 1 GHz. The thick semi-transparent gray line represent the adiabatic limit which is calculated from the DC map Eq.\ (1) in main text.
\textbf{b.} Similar to \textbf{a} but for frequencies ranging from 1 GHz to 17 GHz.
}
\end{figure*}

We interpret the deviations from adiabatic behaviour as the onset of the dynamical regime.
The crossover frequency can be corroborated from analysing the contribution of the different modes of the TCW.
Fig.~S10a shows the propagating modes and Fig.~S10b the transmission in the TCW. 
The contribution of each pair of modes to the transmission is shown in Fig.~S10c. 
One finds that the modes close to the Fermi energy at $E_F = 0$ are responsible for the most non-linear behaviour, giving rise to rectification.
The velocity of these modes can be estimated from $v_i= \sqrt{\frac{2(E-E_i)}{m*}}$ where $E_i$ is the transversal energy of the mode $i$. For the modes represented in red in Fig.~S9a one finds $2.6 \times 10^4$ m\,s$^{-1}$ and $6.3 \times 10^4$ m\,s$^{-1}$ which correspond to propagation times of 77~ps and 34~ps in the TCW, respectively. 
This is of similar order of magnitude as the frequency where we see deviations from the adiabatic regime.

\begin{figure*}[h!]
\centering
\includegraphics[scale=1]{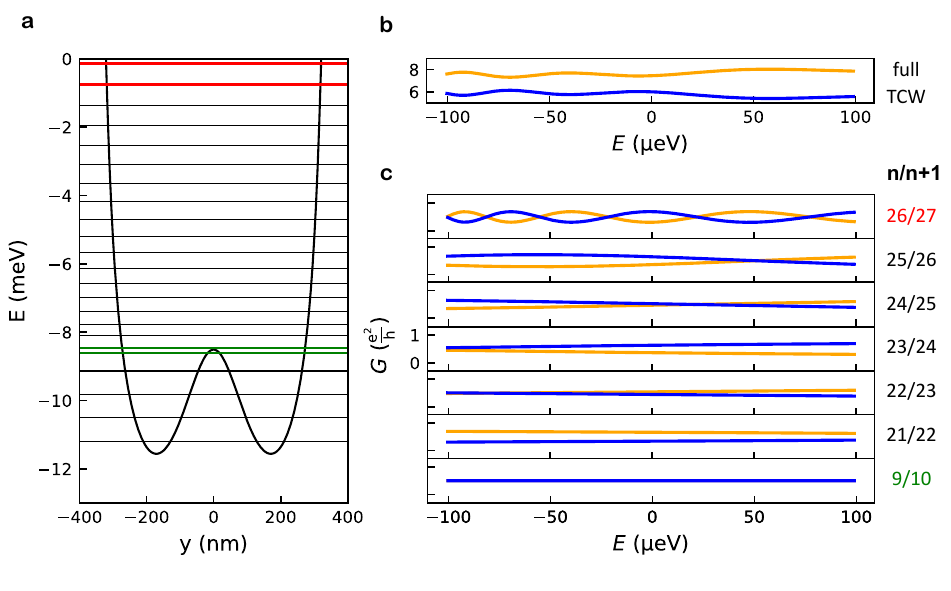}
\caption*{\textbf{Figure~S10: Mode decomposition of the TCW.} \textbf{a.} Energies of the propagating modes in the TCW. The modes that are coupled through the barrier are represented in green. The modes that are the closest to the Fermi energy $E_F$ = 0 are represented in red. \textbf{b.} Total transmission of the TCW as function of the energy. 
\textbf{c.} Decomposition of the total transmission of the TCW into the contribution of each pair of modes. It shows that the energy dependence is mainly due to the modes near the Fermi energy.}
\end{figure*}

\pagebreak
\printbibliography[heading=subbibliography,title={References}]
\end{refsection}

\end{document}